\documentclass[12pt, letter]{article}

\usepackage{geometry}
\usepackage{setspace}
\usepackage{stix}
\usepackage{titling}
\usepackage{authblk}
\usepackage{mathtools}
\usepackage{xcolor, hyperref}
\hypersetup{colorlinks=true, linkcolor=blue, citecolor=blue, urlcolor=cyan}
\usepackage{cite}
\usepackage{algpseudocode} 
\usepackage{algorithmicx}
\usepackage{algorithm}
\usepackage{graphicx}
\usepackage{subcaption}
\usepackage{siunitx}
\usepackage{indentfirst}
\usepackage{pgfgantt}
\usepackage[nottoc,numbib]{tocbibind}
\usepackage{cases}
\usepackage{multirow}

\DeclarePairedDelimiter{\ps}{\lparen}{\rparen}
\DeclarePairedDelimiter{\bs}{\lbrack}{\rbrack}

\DeclarePairedDelimiter{\br}{\{}{\}}

\makeatletter
\newcommand{\pushright}[1]{\ifmeasuring@#1\else\omit$\displaystyle#1$\ignorespaces\fi}
\makeatother

\title{Numerical Model of Thermionic- and Photo- emission Electron Heat Spreading}

\author[1]{Indronil Ghosh\footnote{email: ighosh@g.ucla.edu}}
\author[1]{Timothy S. Fisher}
\affil[1]{University of California, Los Angeles}

\date{ }
\begin{document}
\begin{titlepage}
\maketitle
\begin{abstract}
\noindent Thermionic emission has been exploited to give rise to the theory of thermionic cooling also known as electron transpiration cooling, which can potentially serve as a powerful and engineerable cooling mode for hypersonic leading edges that can reach temperatures exceeding 2000 °C. However, the contribution to this cooling mode by photoexcited electrons remains relatively unexplored. Here, we present a numerical model of thermionic emission and photoemission driven cooling and heat spreading, examining the trajectories of electrons emitted based on a random energy model within a prescribed potential space. By simulating surfaces with two different temperature gradients, and imposing potential spaces derived for Cartesian, cylindrical, and spherical coordinate systems, we demonstrate that heat spreading can be significant for temperature gradients on a length scale comparable to the electron spreading distance. Additionally, by testing two different leading edge radii, we find that heat spreading affects a larger percentage of surface area for a smaller leading edge radius. 

\end{abstract}
\end{titlepage}
\tableofcontents
\newpage
\newgeometry{top=1in,bottom=1in,right=1in,left=1in}
\section{Introduction}
In addition to thermionically driven cooling, here we investigate the contribution of photo-excited electrons to heat spreading. The concept of an electron excited by a photon dates back to Einstein's Noble Prize winning work on the photoelectric effect \cite{Einstein1905}, in which he theorized that the kinetic energy of an electron emitted from a surface is equal to the incident photon energy substracted by the material work function energy. Photoemission theory was also informed by the theory of field emission of electrons by Fowler and Nordheim \cite{FowlerNordheim1928}, in which they derived the electron current due to electric field acting on a metal. Then, Fowler \cite{Fowler1931} and DuBridge \cite{DuBridge1933} motivated probability distributions for the energy of photoemitted electrons, as a function of temperature and incident light frequency. 

Recently, Jensen \cite{Jensen2006a,Jensen2006b} has generalized the Fowler-DuBridge model to the ``modified Fowler-DuBridge photocurrent model'' to apply it to coated metal surfaces, studying the emitted current, quantum efficiency, and thermal evolution of photocathodes with comparison to experiment. Jensen \cite{Jensen2007} expanded upon this model, incorporating electron scattering relaxation time and moments of the momentum distribution, newly studying cesiated surfaces with comparison to experiment. Another theory of photoemission using the alpha-semiconductor model was proposed and used to correspond with experimental data on the quantum efficiency of cesium antimonide \cite{Jensen2008}. His model of photocathode emittance using the moments of distribution function approach has been advanced to account for first-order effects of temperature and electric field \cite{Jensen2010}.

A review of photoemission theory and photocathode development is provided in 2009 \cite{Jensen2009}, which summarizes Spicer and Berglund et al.'s semi-empirical ``Three-Step Model'' \cite{Spicer1958, Berglund1964, Sommer1965, Spicer1993}, in addition to the Fowler-DuBridge model and Jensen's moments-based photoemission theory. The ``Three-Step Model'' is notable for assisting experimentalists in interpreting the steps of an electron's photo-excitation, its motion in the material, and finally its overcoming the work function barrier of the material to escape. 

We note that supplementing thermionic emission with photoemission has been tested to generate electricity using a solar-concentrator system \cite{Schwede2010} and that this process has been further optimized through the use of gallium arsenide heterostructures \cite{Schwede2013}. Finally, in a review on thermionic emission and photoemission for energy-conversion, McCarthy et al. \cite{McCarthy2014} introduced a ``random energy model'' model that captures the electron energy distribution for two-dimensional (2D) materials and separately for three-dimensional (3D) materials. We make use of this random energy model to assess heat spreading driven by the combination of thermionic emission and photoemission. 
\newpage
\section{Methods}
\begin{figure}[H]
    \centering
    \includegraphics[width=1\linewidth]{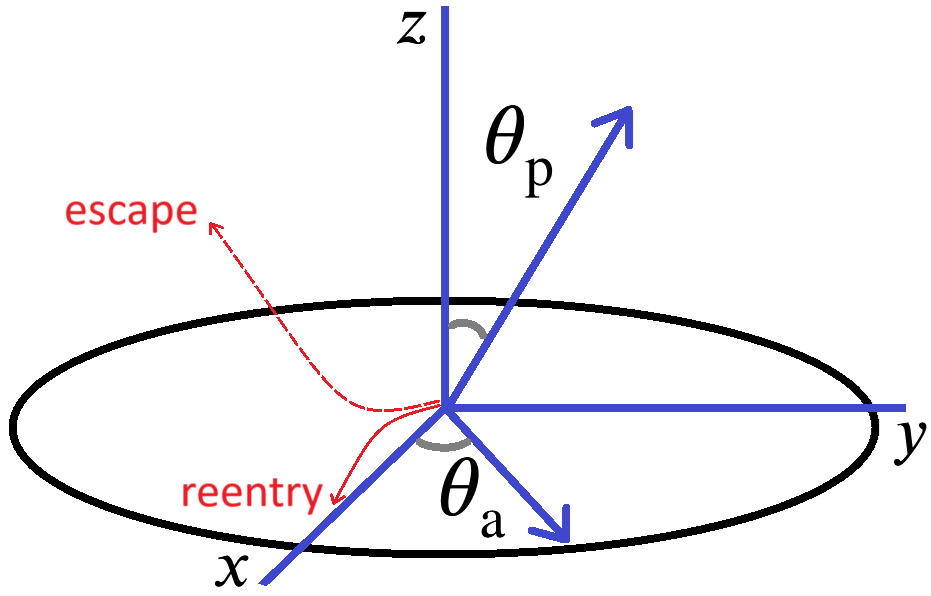}
    \caption{Diagram of sample (black circle) with Cartesian coordinate axes, polar angle \(\theta_\text{p}\), azimuthal angle \(\theta_\text{a}\), as well as illustration of example emitted electron trajectories. Electrons escape the material entirely if they have sufficient kinetic energy to overcome the virtual cathode (VC), and reenter the material surface otherwise.}
    \label{fig:PECoordSys}
\end{figure}

\begin{figure}[H]
    \centering
    \includegraphics[width=1\linewidth]{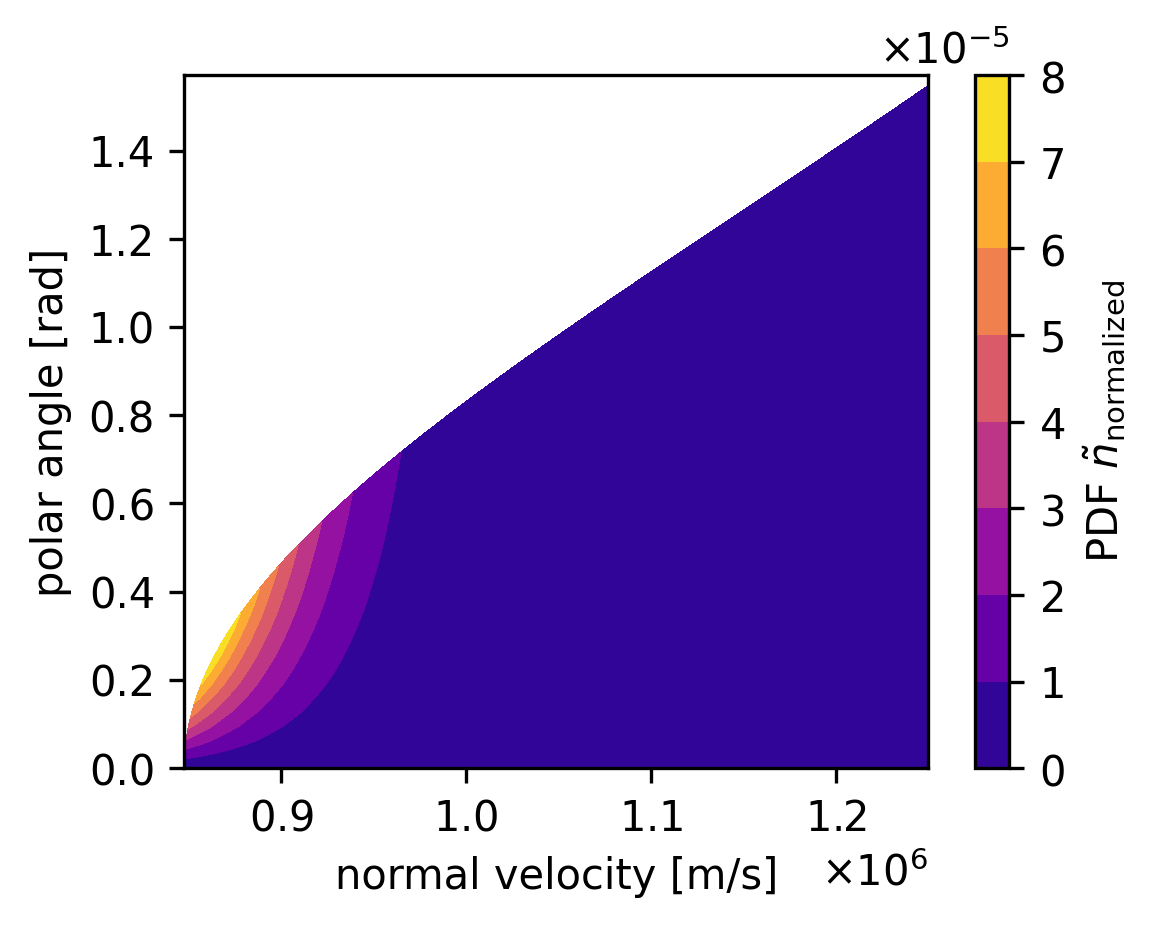}
    \caption{Example electron velocity PDF for surface temperature of \SI{2000}{K}.}
    \label{fig:REMElectronVelocityPDF}
\end{figure}

\begin{table}[H]
\centering
\begin{tabular}{||c | c ||} 
 \hline\hline
 \(u\) [m/s] & normal velocity \\ \hline
 \(\theta_\text{p}\) [rad] & polar angle \\ \hline
 \(\theta_\text{a}\) [rad] & azimuthal angle \\ \hline
 \(r_\text{sample}\) [m] & sample radius \\ \hline
 \(\Delta r\) [m] & mean radial distance between sample mesh points \\ \hline
 \(T\) [K] & temperature \\ \hline
 \(E_\text{F} = 0\) eV & Fermi energy \\ \hline
 \(m_\text{e} = 9.11 \times 10^{-31}\) kg & electron mass \\ \hline
 \(k_b = 1.38\times 10^{-23}\) J/K & Boltzmann constant \\ \hline
 \(\hbar = 1.05\times10^{-34}\) J\(\cdot\)s & reduced Plank constant \\ \hline
 \(\omega\) [rad/s] & light angular frequency \\ \hline
 \(\varphi_\text{work}\) [eV] &  work function energy \\ \hline
 \(\phi_\text{VC}\) [V] & virtual cathode potential \\ \hline
 \(\phi_\text{W}\) [V] & wall potential \\ \hline
  \(\delta()\) & delta function \\[0.5ex]
 \hline\hline
\end{tabular}
\caption{Nomenclature}
\label{table:NomenclaturePhotoemission}
\end{table}
Here we develop a numerical model of thermionic and photoemission driven cooling, to compare against the PIC-based modeling and future photoemission experiments. In essence, the numerical model applies the 3D random energy model of McCarthy et al. \cite{McCarthy2014} to describe the electron energy distribution (EED). The physical system analyzed includes a prescribable potential space between the cathode and anode, temperature gradient along the cathode surface, and lateral electron travel and redeposition calculated by the Runge-Kutta method. The current work limits the incident light to one frequency at the middle of the visible range, but next steps will extend the work to a range of incident light frequencies. The foundation of the random energy model is the probability distribution function (PDF) for electron velocity below, which takes normal velocity \(u\) and polar angle \(\theta_\text{p}\), for a certain temperature \(T\):
\begin{equation}\label{eq:REMVelocityDistFunc}
\tilde{n}\ps*{u,\theta_\text{p}, T}_\text{3D} \, du\,  d\theta_\text{p} = T \ln\bs*{\exp\ps*{\frac{E_\text{F} - \frac{1}{2}m_\text{e} u^2}{k_b T}} + 1} \sin\ps*{\theta_\text{p}} \, du\, d\theta_\text{p}\,.
\end{equation}
Normal velocity \(u\) is defined as the vertical component of electron emission velocity (i.e., in the \(z\) direction in Fig. \ref{fig:PECoordSys}) before an electron loses the material work function energy upon exiting the material, and \(\theta_\text{p}\) is defined as the polar angle from the surface normal (in the \(z\) direction in Fig. \ref{fig:PECoordSys}); these definitions reflect the conventions used by McCarthy et al. \cite{McCarthy2014}. In Fig. \ref{fig:PECoordSys}, we illustrate the coordinate system and angles at an arbitrary site of electron emission. The PDF of Eq. \eqref{eq:REMVelocityDistFunc} is normalized by the number of electrons available for emission, \(N_{\text{avail},\,\Delta \theta_\text{p}\text{-3D}}\), or mathematically the joint CDF evaluated at the upper bounds of the input variables.
\begin{equation}\label{eq:REMNumberOfElectronsAvailable}
N_{\text{avail}, \Delta \theta_\text{p}\text{-3D}}\ps*{T} = \int\limits_0^{\pi/2}\int\limits_{u_\text{min} = \sqrt{2\bs*{E_\text{F}+\varphi_\text{work} - \hbar \omega \cos\ps*{\theta_\text{p}}}/m_\text{e}} }^\infty T \ln\bs*{\exp\ps*{\frac{E_\text{F} - \frac{1}{2}m_\text{e} u^2}{k_b T}} + 1} \sin\ps*{\theta_\text{p}} \, du\, d\theta_\text{p}
\end{equation}
\begin{equation}\label{eq:REMVelocityDistNormalized}
\tilde{n}_\text{normalized}\ps*{u, \theta_\text{p}, T} = \tilde{n} \ps*{u, \theta_\text{p}, T}_\text{3D}/N_{\text{avail}, \Delta \theta_\text{p} \text{-3D}}\ps*{T}
\end{equation}
The PDF can be visualized as a two-dimensional contour map in Fig. \ref{fig:REMElectronVelocityPDF}, illustrating that emission of electrons at low polar angle and low normal velocity is most likely. The white region of the graph indicates that electrons with too low of a normal velocity at a high polar angle will not successfully emit, having insufficient kinetic energy (i.e., \(u < u_\text{min}\) as per Eq. \eqref{eq:REMNumberOfElectronsAvailable}) to escape the work function barrier of the material.

\begin{figure}[H]
\centering
\begin{subfigure}{.49\textwidth}
    \centering
    \includegraphics[width=1\textwidth]{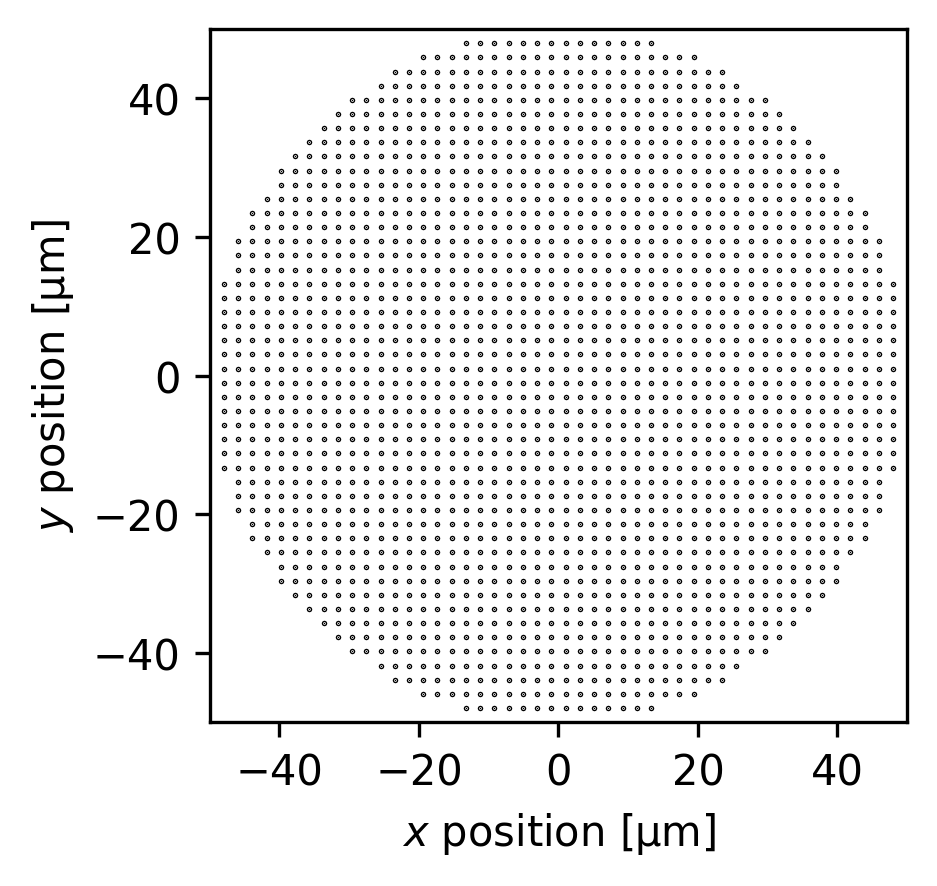}
    \caption{Point mesh representing sample surface}
    \label{fig:SpreadingMeshFull}
\end{subfigure}
\begin{subfigure}{.49\textwidth}
    \centering
    \includegraphics[width=1\textwidth]{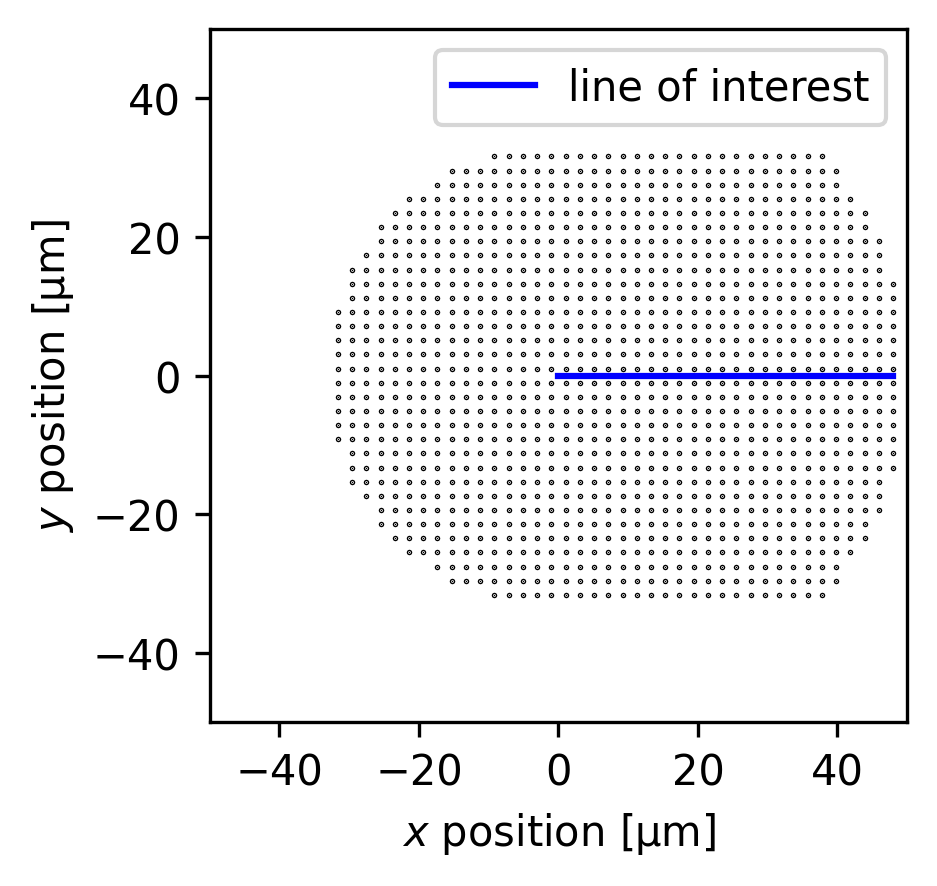}
    \caption{Reduced mesh and line of points studied given azimuthal rotational symmetry}
    \label{fig:SpreadingReducedMesh}
\end{subfigure}
\caption{Meshes for numerical model of thermionic and photoemission heat spreading, where \(x\) and \(y\) positions denote positions in the plane of the sample surface, which can also be located with radial position \(r\) and azimuthal angle \(\theta_\text{a}\)}
\label{fig:NumericalREMModelMeshes}
\end{figure}

The reflected current distribution as a function of normal velocity and polar angle is developed by considering that electrons reflect if they escape the work function barrier of the material, but cannot overcome the virtual cathode of the potential space. The following Eqs. \eqref{eq:PEEDReflected}-\eqref{eq:PEEDWorkPlusGap} encapsulate this description of the reflected current.

\begin{align}
& J_\text{PEED}^\text{work}\ps*{u, \theta_\text{p}, T}\, du\,d\theta_\text{p} = \nonumber \\
& \frac{m_\text{e}^2}{2 \pi^2 \hbar^3} \frac{u\ps*{\frac{1}{2}m_\text{e} u^2 + \hbar \omega \cos \theta_\text{p} - E_F - \varphi_\text{work}}^2\;H\ps*{\frac{1}{2}m_\text{e} u^2 + \hbar \omega \cos \theta_\text{p} - E_F - \varphi_\text{work}}}{1 + \exp\ps*{\frac{\frac{1}{2}m_\text{e} u^2 - E_F}{k_B T}}}\, du\,d\theta_\text{p}\label{eq:PEEDWork}\\
& J_\text{PEED}^\text{work+VC}\ps*{u, \theta_\text{p}, T}\, du\,d\theta_\text{p} = \nonumber\\ 
&\frac{m_\text{e}^2}{2 \pi^2 \hbar^3} \frac{u\ps*{\frac{1}{2}m_\text{e} u^2 + \hbar \omega \cos \theta_\text{p} - E_F - \varphi_\text{work}}^2}{1 + \exp\ps*{\frac{\frac{1}{2}m_\text{e} u^2 - E_F}{k_B T}}} \nonumber \\
& \cdot H\bs*{\frac{1}{2}m_\text{e} u^2 + \hbar \omega \cos \theta_\text{p} - E_F - \varphi_\text{work} + q_\text{e}\ps*{\phi_\text{VC} - \phi_\text{W}}} \, du\,d\theta_\text{p}\label{eq:PEEDWorkPlusGap}\\
& J_\text{PEED}^\text{reflected}\ps*{u, \theta_\text{p}, T}\, du\,d\theta_\text{p} = \tilde{n}_\text{normalized}\ps*{u, \theta_\text{p}, T} \;  \bs*{J_\text{PEED}^\text{work}\ps*{u, \theta_\text{p}, T} - J_\text{PEED}^\text{work+gap}\ps*{u, \theta_\text{p}, T}}\, du\,d\theta_\text{p}\label{eq:PEEDReflected}
\end{align}
Employing the methods behind the sheath potential models for Cartesian, cylindrical, and spherical coordinate systems from \cite{Ghosh2025}, we prescribe the following three potential spaces to be the sheathes extending normal from the sample in the results that follow. 
\begin{figure}[H]
\centering
\begin{subfigure}{.3\textwidth}
    \centering
    \includegraphics[width=1\linewidth]{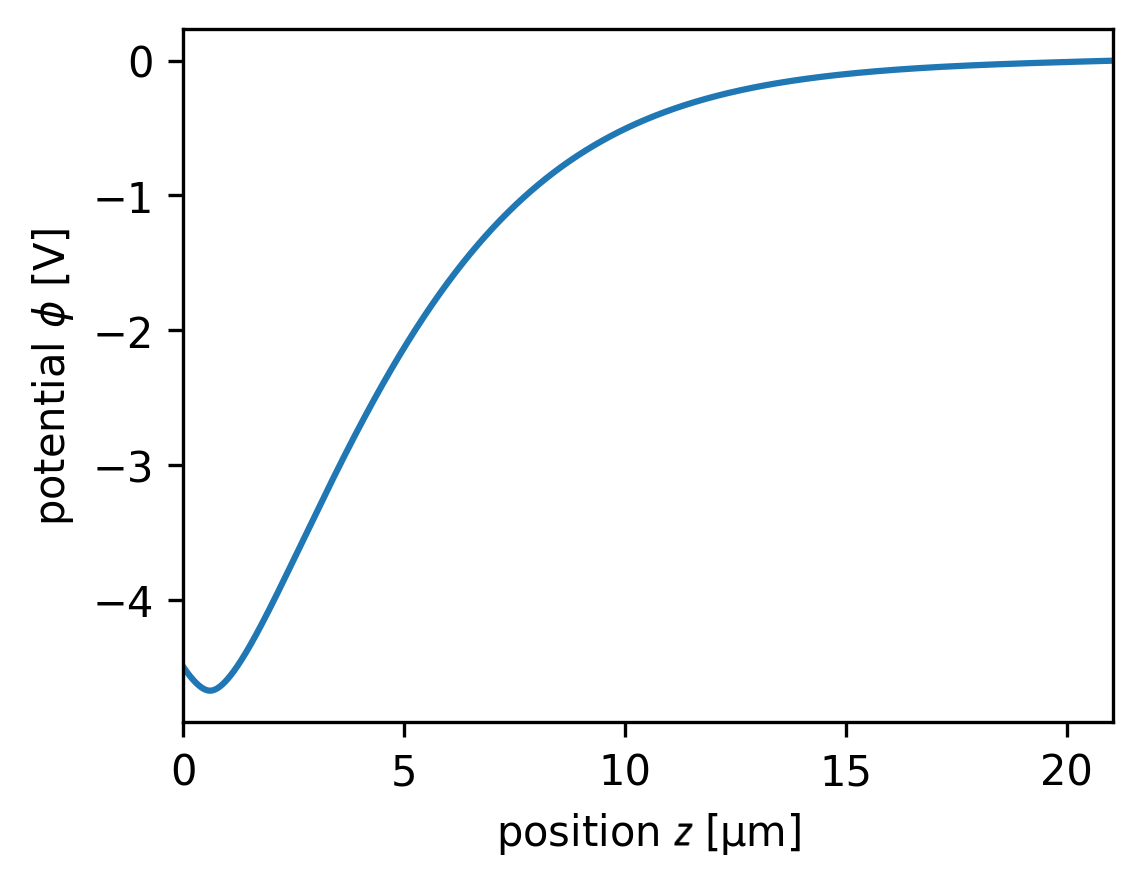}
    \caption{Cartesian, \(\phi_\text{VC} -\phi_\text{W} = \SI{-0.175}{V}\)}
    \label{fig:CarPotential}
\end{subfigure}
\hfill
\begin{subfigure}{.3\textwidth}
    \centering
    \includegraphics[width=1\linewidth]{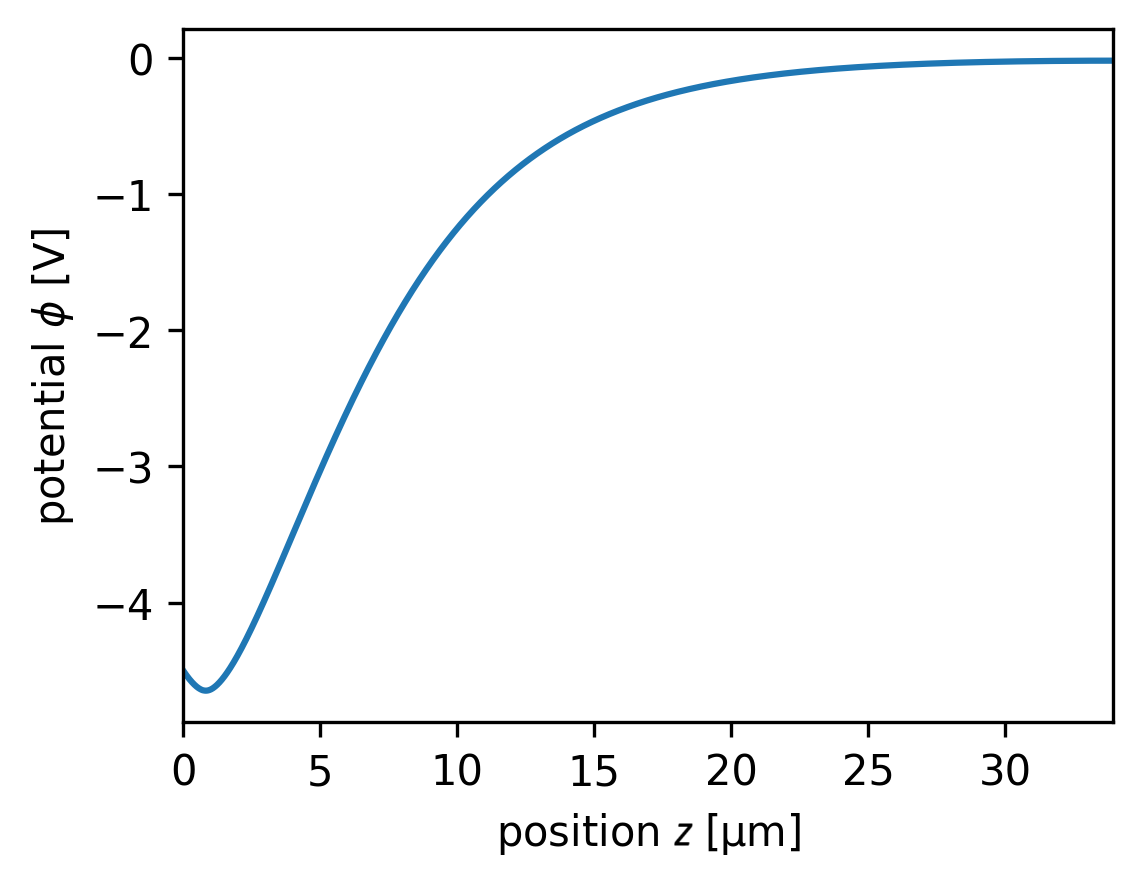}
    \caption{cylindrical, \(\phi_\text{VC} -\phi_\text{W} = \SI{-0.146}{V}\)}
    \label{fig:CylPotential}
\end{subfigure}
\hfill
\begin{subfigure}{.3\textwidth}
    \centering
    \includegraphics[width=1\linewidth]{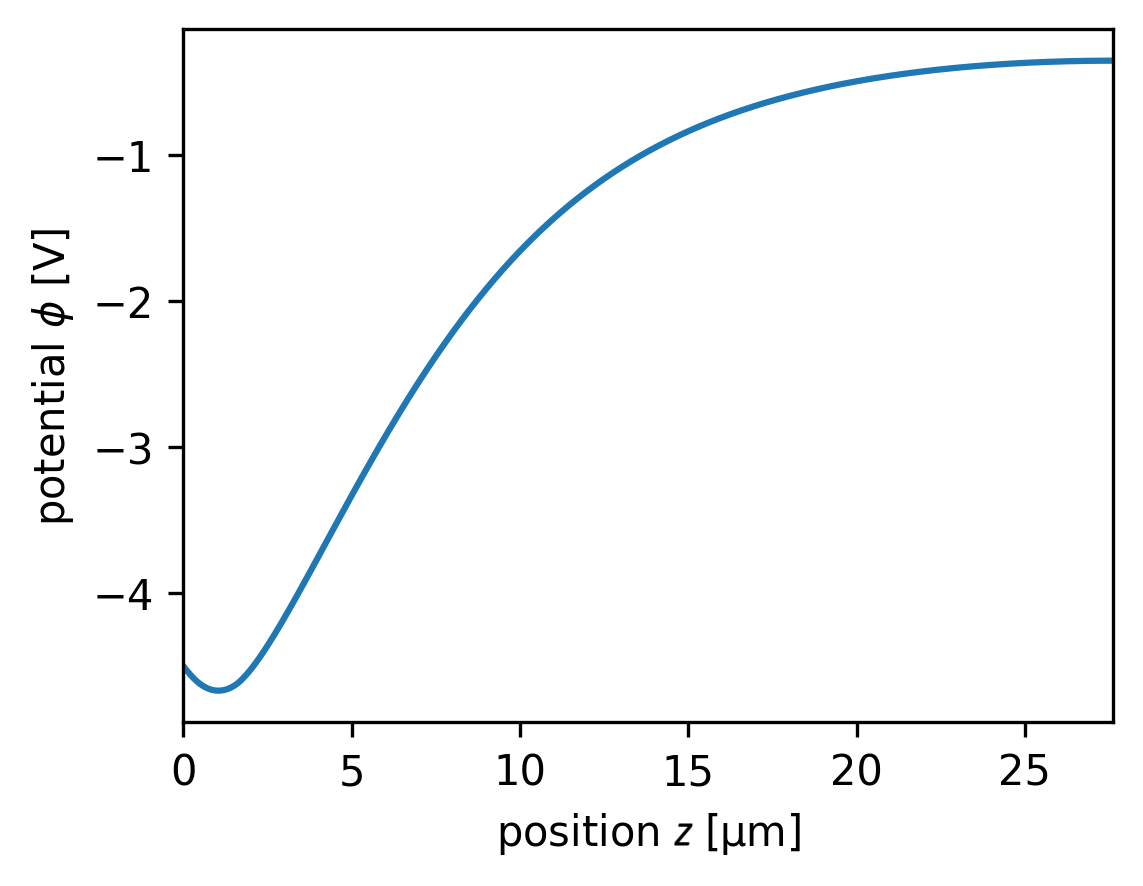}
    \caption{spherical, \(\phi_\text{VC} -\phi_\text{W} = \SI{-0.168}{V}\)}
    \label{fig:SphPotential}
\end{subfigure}
\caption{Prescribed potential spaces for each of three coordinate systems, where position \(z\) is normal to the surface following the convention of Fig. \ref{fig:PECoordSys}.}
\label{fig:PotentialSpaces}
\end{figure}
To generate these potential spaces using our generalized methodology, we assume for our first parametric study a light angular frequency of \(\omega = 3.73 \times 10^{15}\) rad/s, work function energy of \(\varphi_\text{work} = \SI{4.5}{eV}\), and leading edge (LE) radius of \SI{1}{\centi\meter}, which constrains the quasi-neutral plasma density to \(2.42 \times 10^{19}\;\text{m}^{-3}\). To compare, we provide a second parametric study, reducing the LE radius to \SI{1}{\milli\meter}, which constrains the quasi-neutral plasma density to  \(2.42 \times 10^{21}\;\text{m}^{-3}\). We also aim to maintain similar virtual cathode potentials \(\phi_\text{VC}\) throughout the different spaces, to focus on the effects of other sheath differences, e.g., varying sheath sizes and virtual cathode shapes. To understand how far the electrons travel, we multiply the normal velocity by the tangent of polar angle and by the time before reentry back into the emitter material; consider that this calculation is effectively the lateral velocity component of an electron multiplied by its time spent on a projectile trajectory outside the material. 
\begin{figure}[H]
    \centering
    \includegraphics[width=1\linewidth]{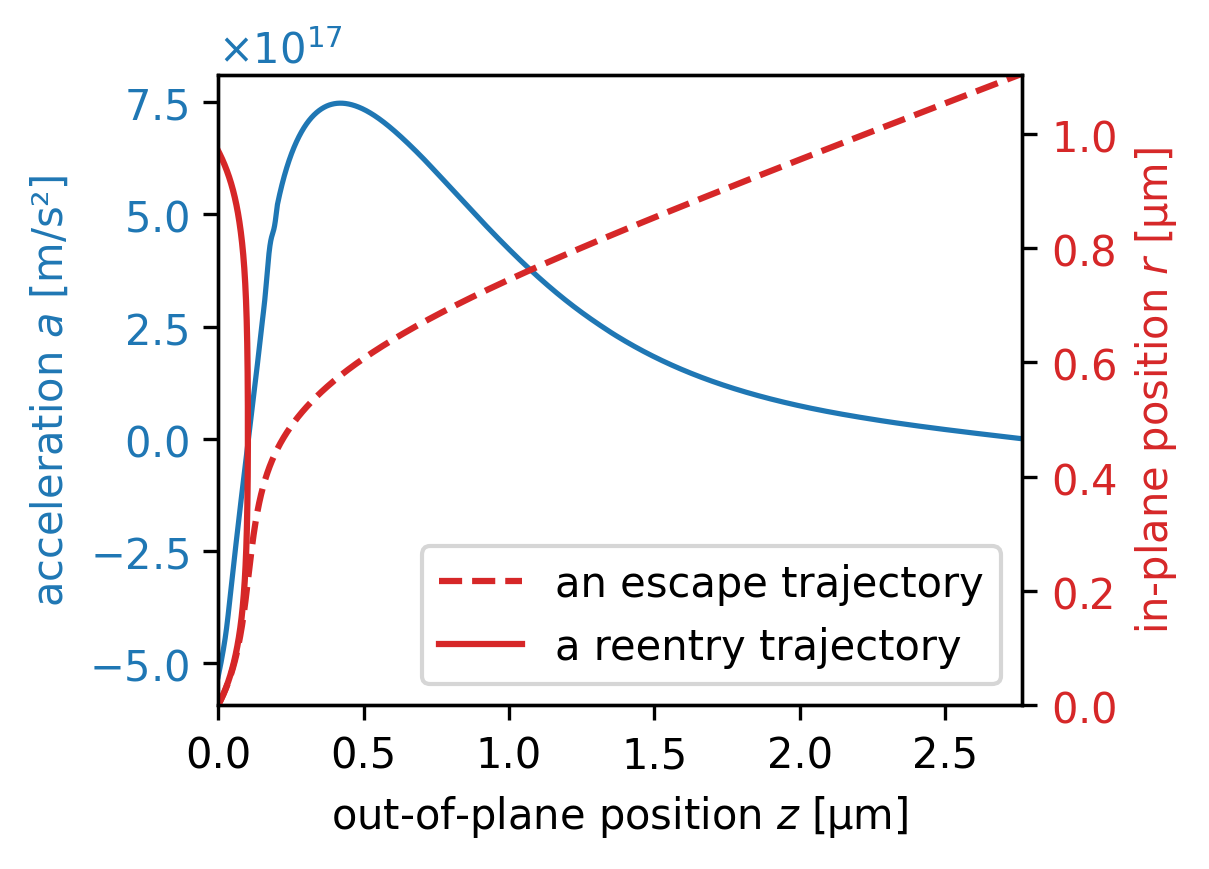}
    \caption{(left \(y\)-axis) Out-of-plane acceleration for the spherical coordinates potential, and (right \(y\)-axis) example trajectories for polar angle \(\theta_\text{p} = 15^\circ\), normal velocity \(u = 9.00\times10^5\) m/s for the escape trajectory, and \(u = 8.98\times10^5\) m/s for the reentry trajectory.}
    \label{fig:AccelTraject}
\end{figure}
The time before reentry is found by applying the Runge-Kutta method of order 5(4) (\verb|scipy.integrate.solve_ivp| \cite{Virtanen2020}) to the first order system of differential equations \( \bs*{\dot{z},\;\text{acceleration}\ps*{z}} \), to calculate the trajectory of an electron normal to the sample surface, when accelerated by a potential space. For example, as calculated in Fig. \ref{fig:AccelTraject} above, an electron that successfully emits from a point on the sample and enters the potential sheath, may either overcome the virtual cathode barrier and escape entirely or experience enough deceleration due to the virtual cathode and reenter the surface. We also artistically render these example trajectories in Fig. \ref{fig:PECoordSys} for electrons emitting from the sample center, and remind that electrons can emit with such trajectories from any \(\ps*{x,\,y}\) or \(\ps*{r,\,\theta_\text{a}}\) position on the sample. Fig. \ref{fig:REMElectronDistance} below illustrates the spaces of electron travel distances for the spherical coordinates potential space, and the aforementioned LE radii. Intuitively, we notice that electrons travel the farthest distance before reentry at high normal velocity and high polar angle, although such electrons also have the lowest probability of emitting as per Fig. \ref{fig:REMElectronVelocityPDF}. While the shape of the travel distance space remains similar for both LE radii, the \SI{1}{\milli\meter} LE radius gives travel distances 1 order of magnitude smaller than those of the \SI{1}{\centi\meter} LE. 
\begin{figure}[H]
\centering
\begin{subfigure}{.49\textwidth}
    \centering
    \includegraphics[width=1\textwidth]{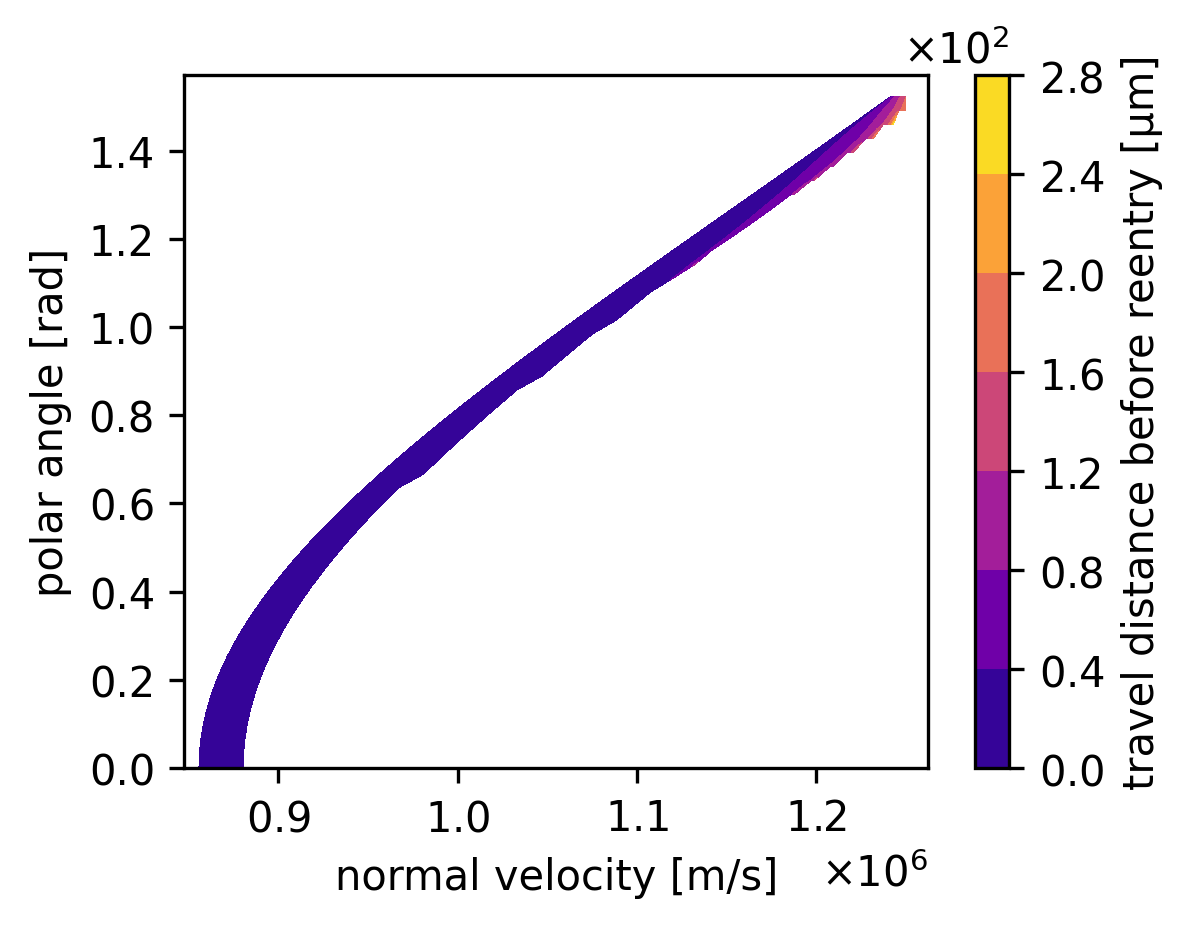}
    \caption{LE radius = \SI{1}{\centi\meter}, mean travel distance = \SI{6.3}{\micro\meter}}
    \label{fig:REMElectronDistance1cm}
\end{subfigure}
\begin{subfigure}{.49\textwidth}
    \centering
    \includegraphics[width=1\textwidth]{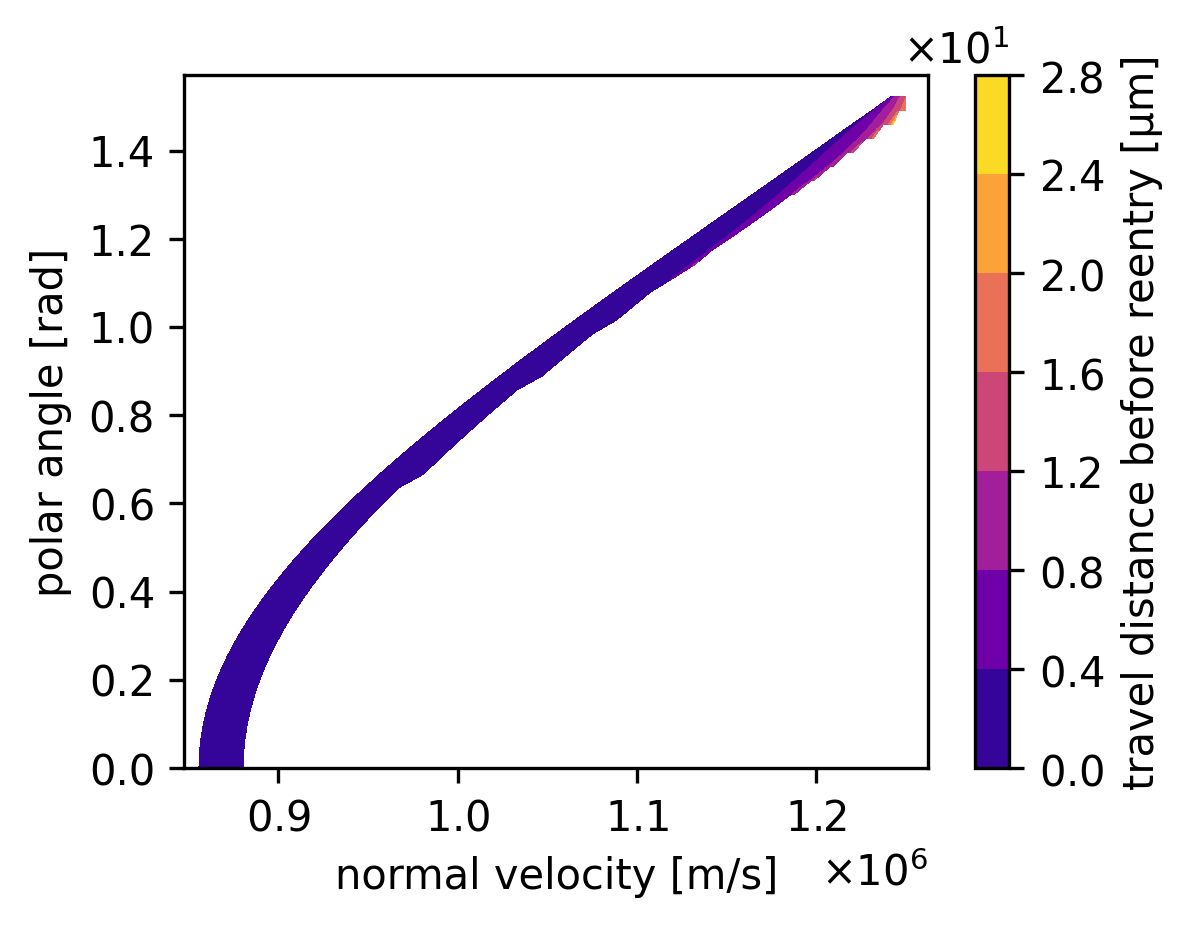}
    \caption{LE radius \SI{1}{\milli\meter}, mean travel distance = \SI{0.63}{\micro\meter}}
    \label{fig:REMElectronDistance1mm}
\end{subfigure}
\caption{Electron travel distances the spherical coordinates potential space, for surface temperature of \SI{2000}{K}.}
\label{fig:REMElectronDistance}
\end{figure}
A 2D circular mesh of points is constructed as in Fig. \ref{fig:SpreadingMeshFull}, i.e., examining a sample surface shaped like a disk with radius \(r_\text{sample} =\) \SI{50}{\micro\meter} where the center of the disk represents the stagnation point of the LE. We wish to determine what reflected heat flux each point receives from all of the other points on the mesh. However, we need not examine every single mesh point and compute the amount of reflected heat flux that it receives, because we can safely assume that the reflected heat flux follows the underlying azimuthmally symmetric temperature gradient of the sample and imply that heat spreading is an isotropic phenomenon for an observer at the origin looking radially outward (again, for our prescribed temperature distributions). By asserting such azimuthal rotational symmetry, we can reduce the mesh under consideration to a set of points close enough to spread electrons to \(y = 0\), between \(x = 0\) and the edge of the sample \(x = r_\text{sample}\) as colored blue in Fig. \ref{fig:SpreadingReducedMesh}. Effectively, a cutoff spreading distance with respect to the points on the line of interest is set, due to the azimuthal symmetry of the underlying temperature gradient. Also, as the deaccelerating potential space yields a mean electron travel distance, it requires that we utilize a mesh sufficiently larger than the mean travel distance to observe the effect of heat spreading due to redepositing electrons. While in reality cylindrically or spherically shaped samples have curved surfaces, this planar mesh is a suitable approximation as the LE radius assumed in deriving the potential spaces is two orders of magnitude larger than the relevant spreading distance.
Below we write equations to distinguish calculation of reflected heat fluxes assuming electron heat spreading (denoted with superscript ``spreading'') and without heat spreading (denoted with superscript ``w/o spread''); together, these two separate approaches yield a comparison of reflected heat fluxes in the case that lateral electron movement is considered and the case that it is not considered. In what we assert to be the more natural and physical case, the calculation of reflected heat flux incorporates contributions from all neighboring points close enough to a point of interest along the line \(y = 0\) in Fig. \ref{fig:NumericalREMModelMeshes}. The delta function \(\delta\) filters for electrons that, from the spectral distribution and prescribed potential space, would travel the distance (denoted as ``dist.'' in \(J_\text{reflected}^\text{spreading}\)) between a neighboring point contributing heat flux and the point of interest. Additionally, the calculation assumes that a neighboring point may emit electrons in any angular direction, hence we must account for the ratio of a mesh point’s area to the area of the annulus where the heat is redeposited (expressed in the final term multiplied in \(J_\text{reflected}^\text{spreading}\)). On the other hand, in the case without heat spreading, we do not consider heat flux contributions from neighboring points, instead asserting that a point of interest along \(y = 0\) only reflects back to itself.
\begin{equation}\label{eq:SpreadingFluxReflected}
	\begin{split}
		J_\text{reflected}^\text{spreading} \ps*{r, \theta_\text{a}} = &\\
        \sum_{i}^{\text{\# mesh points}} & \int_0^\frac{\pi}{2} \int_{u_\text{min}}^{\infty} J_\text{PEED}^\text{reflected}\bs*{u, \theta_\text{p}, T\ps*{r_i, \theta_\text{a}^i}}\\
		& \times \delta\br*{\text{dist. travel}\bs*{u, \theta_\text{p}, T\ps*{r_i, \theta_\text{a}^i}} - \text{dist. between}\bs*{(r, \theta_\text{a}), (r_i, \theta_\text{a}^i)}}\, du\,d\theta_\text{p}\\
		& \times \frac{\text{point area}}{2 \pi \times \text{dist. between}\bs*{(r, \theta_\text{a}), (r_i, \theta_\text{a}^i)} \times \Delta r}
	\end{split}
\end{equation}
\begin{equation}\label{eq:NonspreadingFluxReflected}
	J_\text{reflected}^\text{w/o spread} \ps*{r, \theta_\text{a}} = \int_0^\frac{\pi}{2} \int_{u_\text{min}}^{\infty} J_\text{PEED}^\text{reflected}\bs*{u, \theta_\text{p}, T\ps*{r, \theta_\text{a}}} \, du\,d\theta_\text{p}
\end{equation}
\section{Results \& Discussion}
\begin{figure}[H]
\centering
\begin{subfigure}{.49\textwidth}
    \centering
    \includegraphics[width=\textwidth]{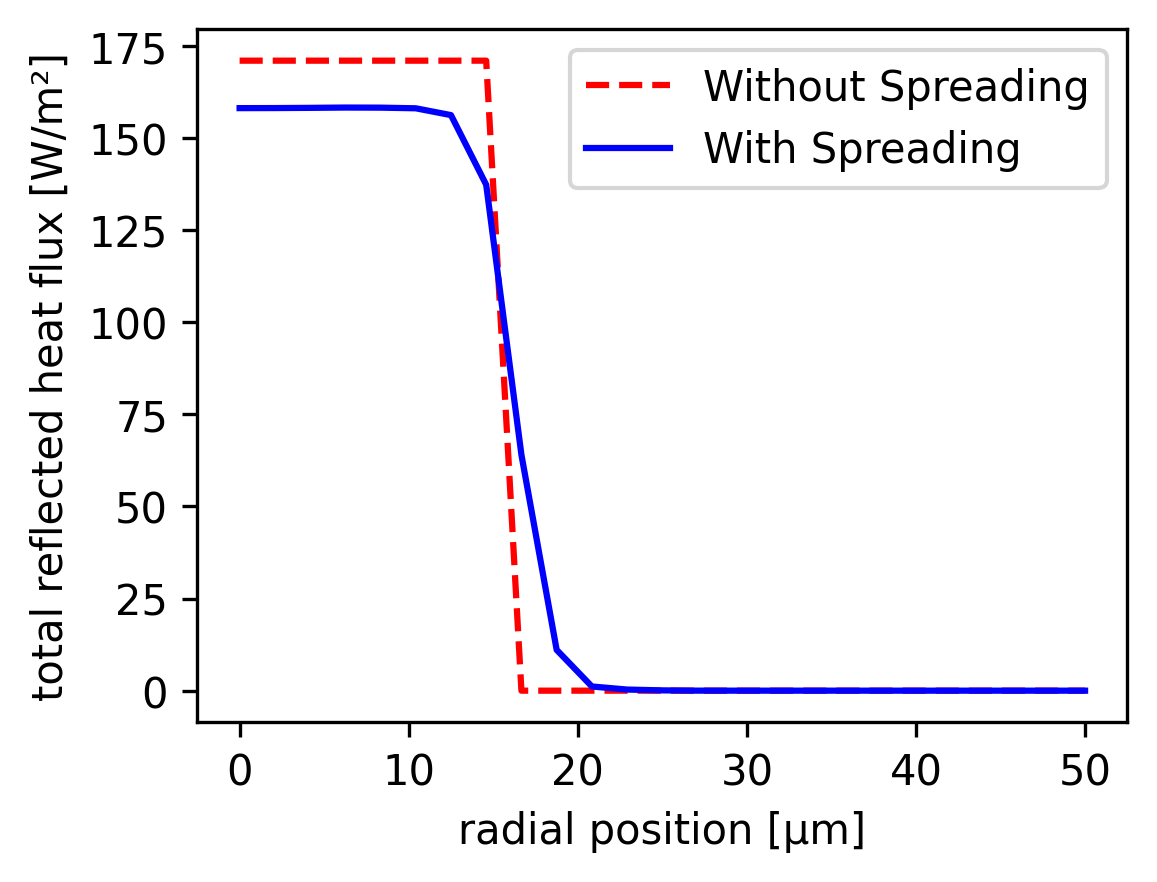}
    \caption{for step function gradient, w/o spread is 8.1\% greater at \(r=0\)}
    \label{fig:CarReflectedHeatFluxStepFunc_1cm}
\end{subfigure}%
\begin{subfigure}{.49\textwidth}
    \centering
    \includegraphics[width=\textwidth]{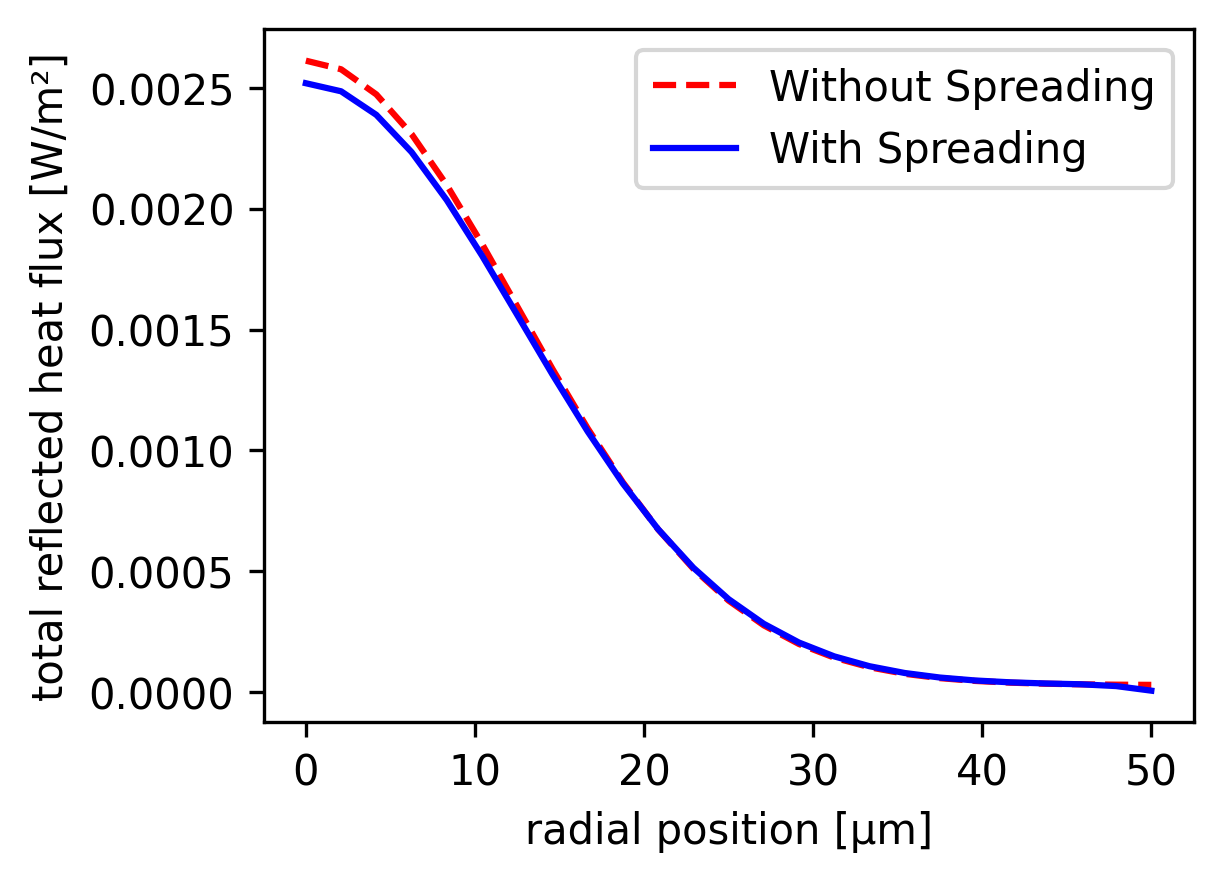}
    \caption{for Gaussian gradient, w/o spread is 3.7\% greater at \(r=0\)}
    \label{fig:CarReflectedHeatFluxGaussian_1cm}
\end{subfigure}
\caption{Total reflected heat fluxes with spreading and without spreading for the Cartesian coordinate system potential space, leading edge radius of \SI{1}{\centi\meter}, and quasi-neutral plasma density of \(2.42 \times 10^{19}\;\text{m}^{-3}\).}
\label{fig:CarREMModelReflectedHeatFluxes_1cm}
\end{figure}

\begin{figure}[H]
\centering
\begin{subfigure}{.49\textwidth}
    \centering
    \includegraphics[width=\textwidth]{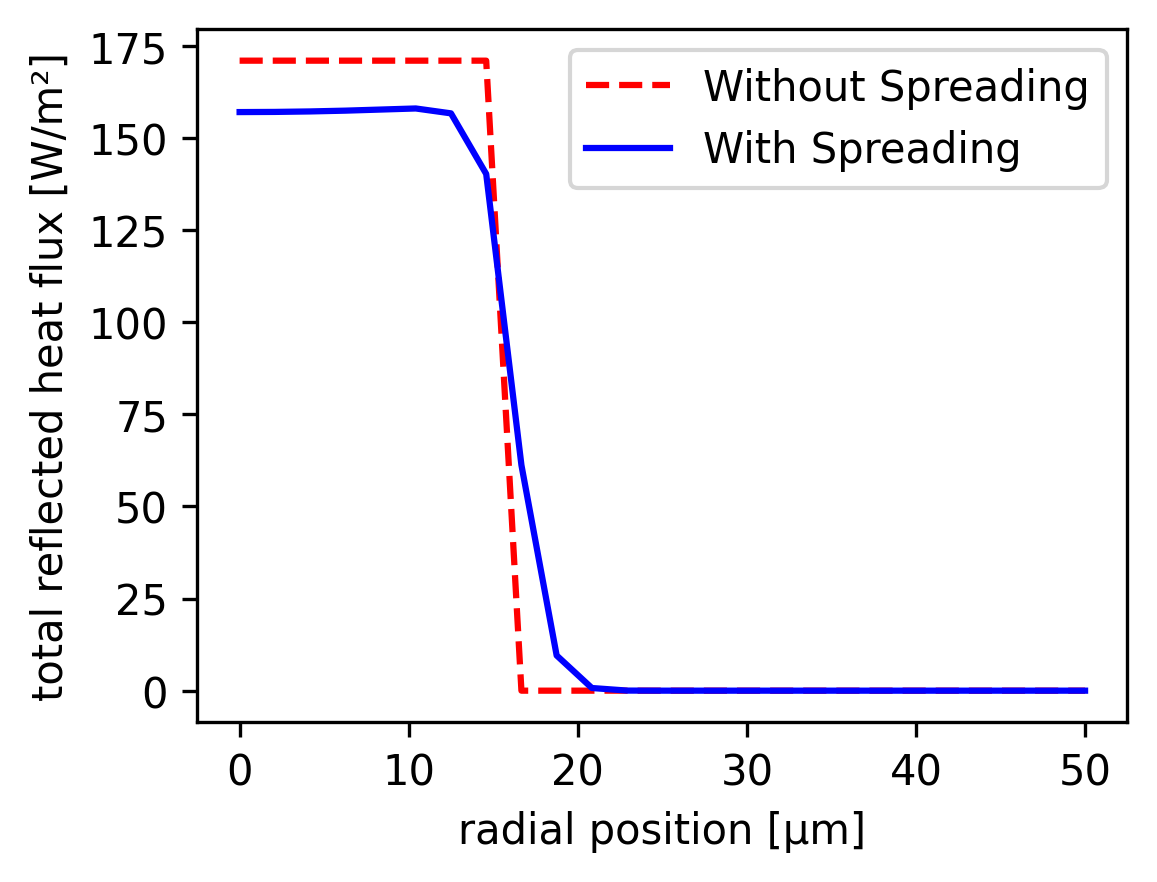}
    \caption{for step function gradient, w/o spread is 8.9\% greater at \(r=0\)}
    \label{fig:CarReflectedHeatFluxStepFunc_1mm}
\end{subfigure}%
\begin{subfigure}{.49\textwidth}
    \centering
    \includegraphics[width=\textwidth]{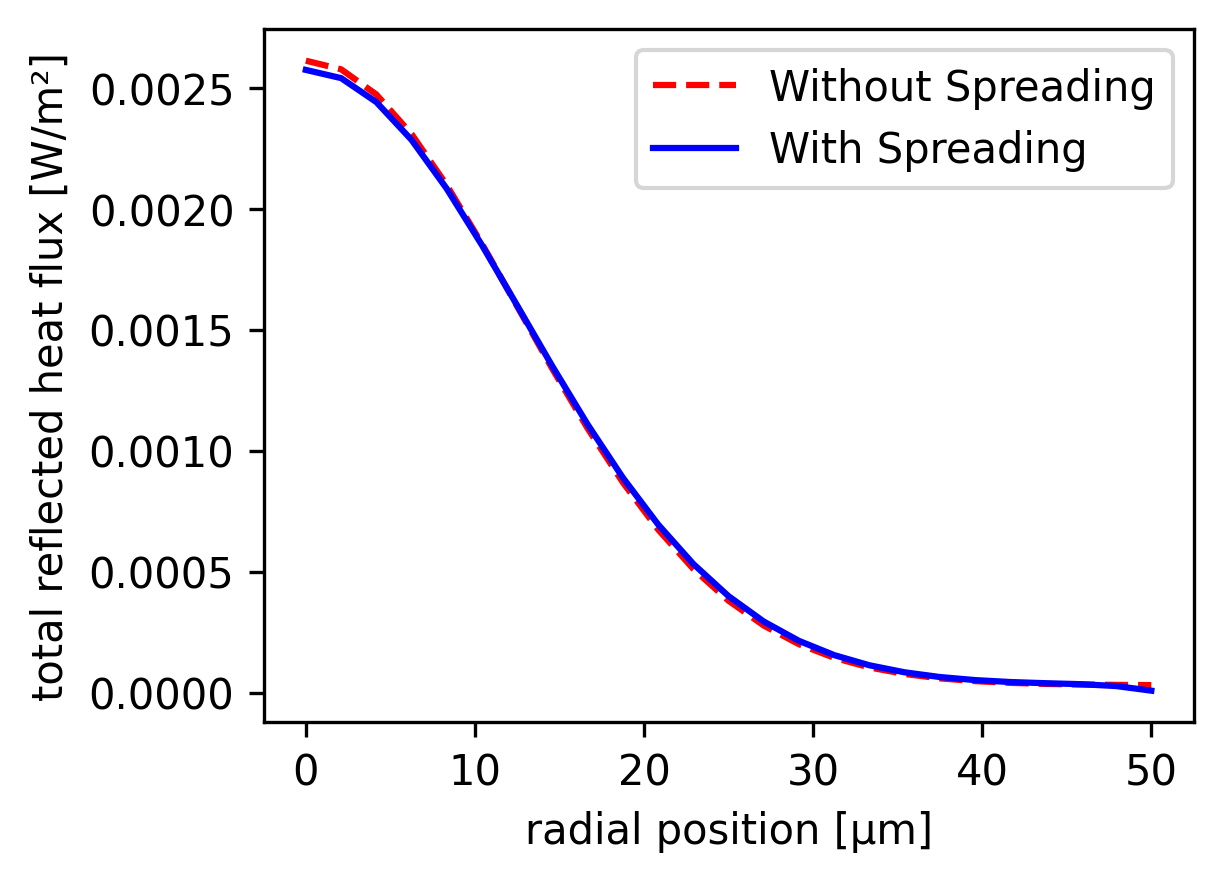}
    \caption{for Gaussian gradient, w/o spread is 1.5\% greater at \(r=0\)}
    \label{fig:CarReflectedHeatFluxGaussian_1mm}
\end{subfigure}
\caption{Total reflected heat fluxes with spreading and without spreading for the Cartesian coordinate system potential space, leading edge radius of \SI{1}{\milli\meter}, and quasi-neutral plasma density of \(2.42 \times 10^{21}\;\text{m}^{-3}\).}
\label{fig:CarREMModelReflectedHeatFluxes_1mm}
\end{figure}

\begin{figure}[H]
\centering
\begin{subfigure}{.49\textwidth}
    \centering
    \includegraphics[width=\textwidth]{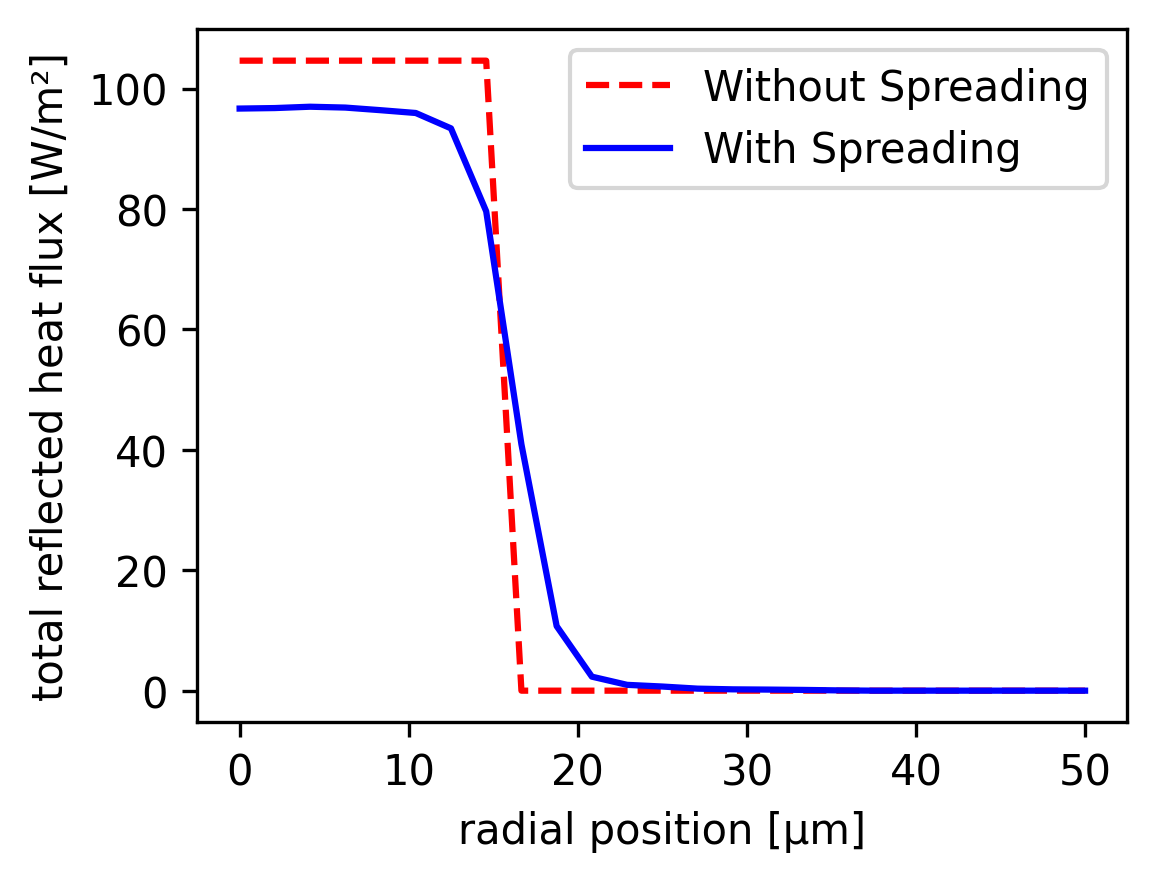}
    \caption{for step function gradient, w/o spread is 8.2\% greater at \(r=0\)}
    \label{fig:CylReflectedHeatFluxStepFunc_1cm}
\end{subfigure}%
\begin{subfigure}{.49\textwidth}
    \centering
    \includegraphics[width=\textwidth]{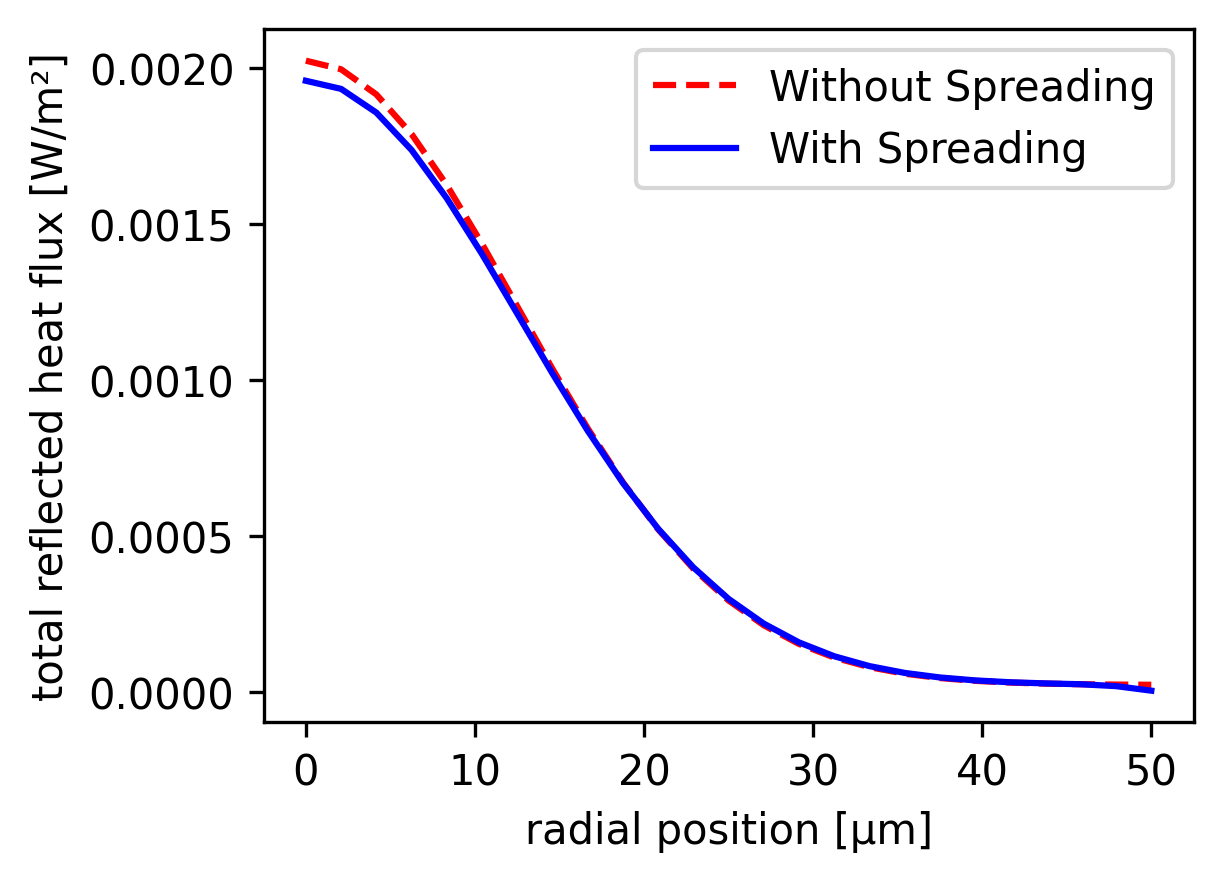}
    \caption{for Gaussian gradient, w/o spread is 3.3\% greater at \(r=0\)}
    \label{fig:CylReflectedHeatFluxGaussian_1cm}
\end{subfigure}
\caption{Total reflected heat fluxes with spreading and without spreading for the cylindrical coordinate system potential space, leading edge radius of \SI{1}{\centi\meter}, and quasi-neutral plasma density of \(2.42 \times 10^{19}\;\text{m}^{-3}\).}
\label{fig:CylREMModelReflectedHeatFluxes_1cm}
\end{figure}

\begin{figure}[H]
\centering
\begin{subfigure}{.49\textwidth}
    \centering
    \includegraphics[width=\textwidth]{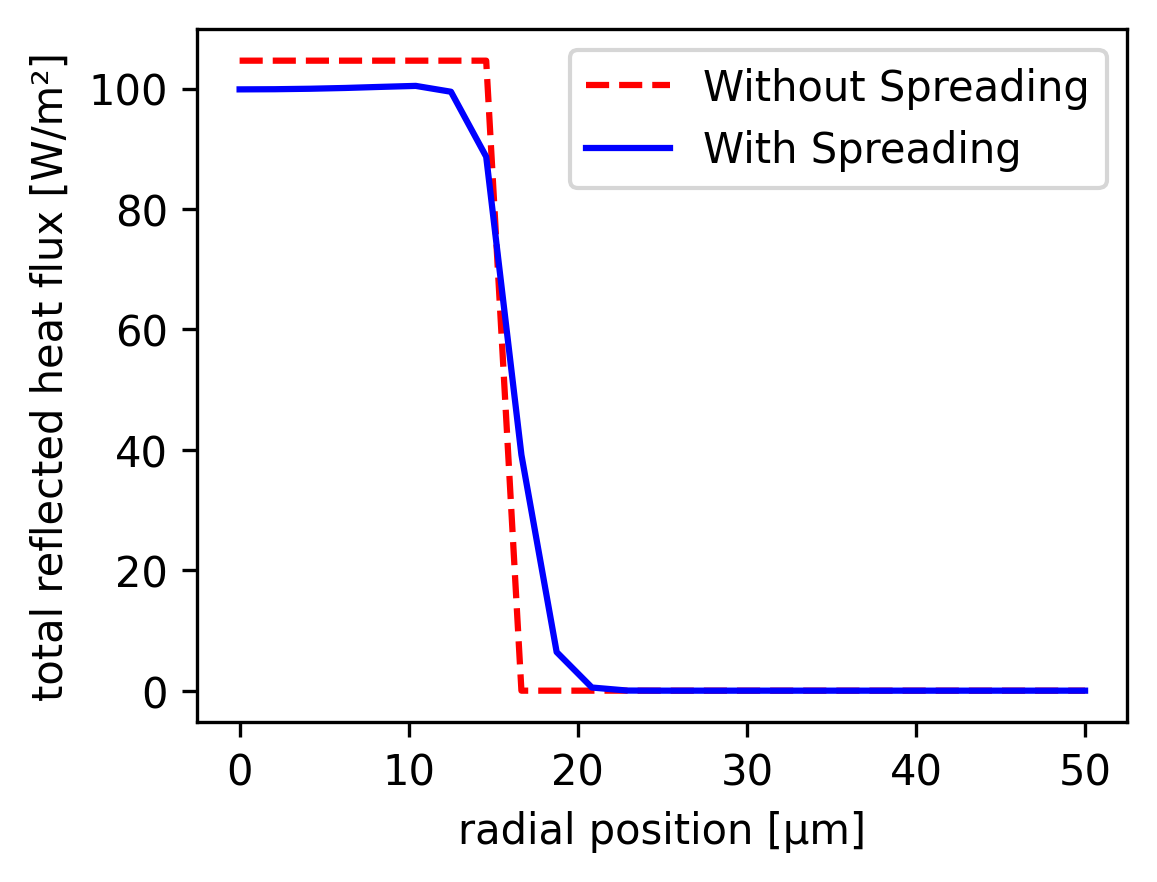}
    \caption{for step function gradient, w/o spread is 4.8\% greater at \(r=0\)}
    \label{fig:CylReflectedHeatFluxStepFunc_1mm}
\end{subfigure}
\begin{subfigure}{.49\textwidth}
    \centering
    \includegraphics[width=\textwidth]{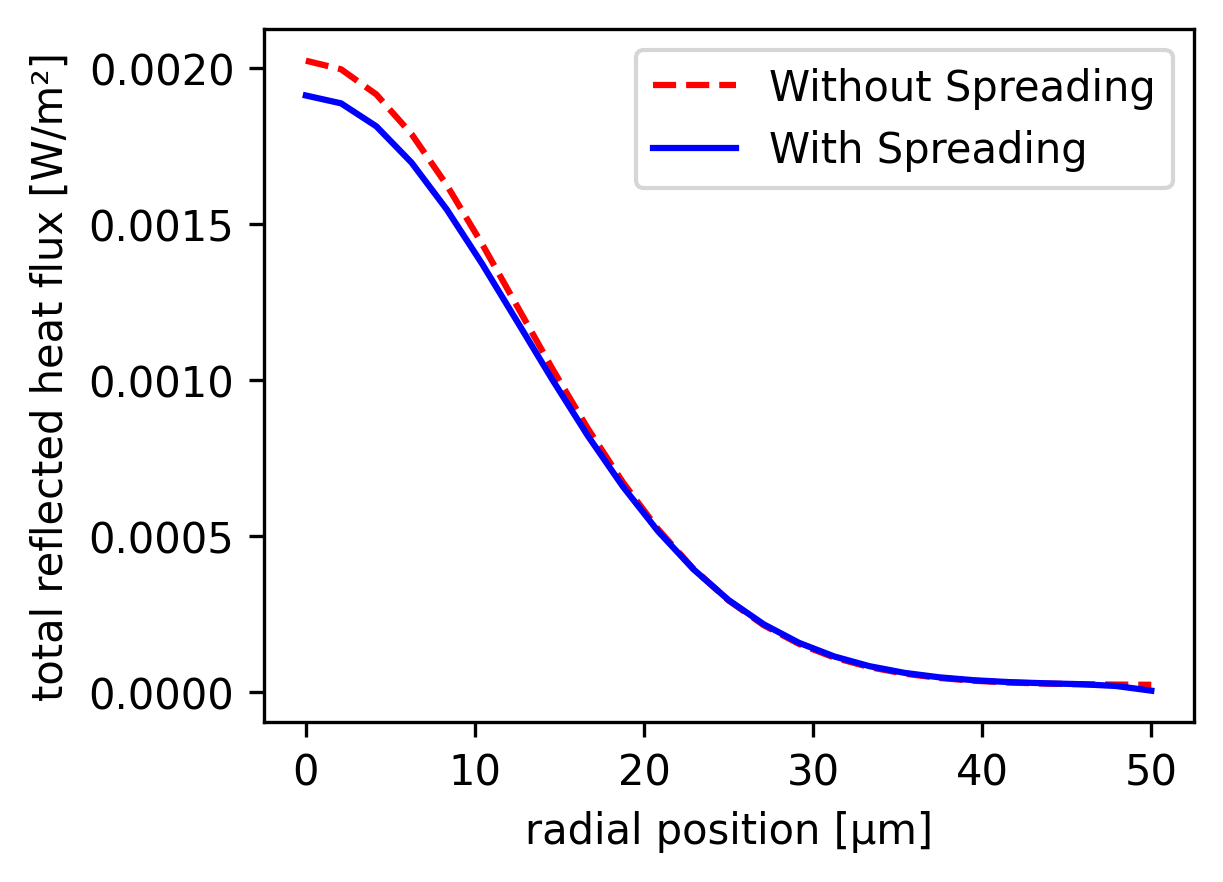}
    \caption{for Gaussian gradient, w/o spread is 5.8\% greater at \(r=0\)}
    \label{fig:CylReflectedHeatFluxGaussian_1mm}
\end{subfigure}
\caption{Total reflected heat fluxes with spreading and without spreading for the cylindrical coordinate system potential space, leading edge radius of \SI{1}{\milli\meter}, and quasi-neutral plasma density of \(2.42 \times 10^{21}\;\text{m}^{-3}\).}
\label{fig:CylREMModelReflectedHeatFluxes_1mm}
\end{figure}

\begin{figure}[H]
\centering
\begin{subfigure}{.49\textwidth}
    \centering
    \includegraphics[width=\textwidth]{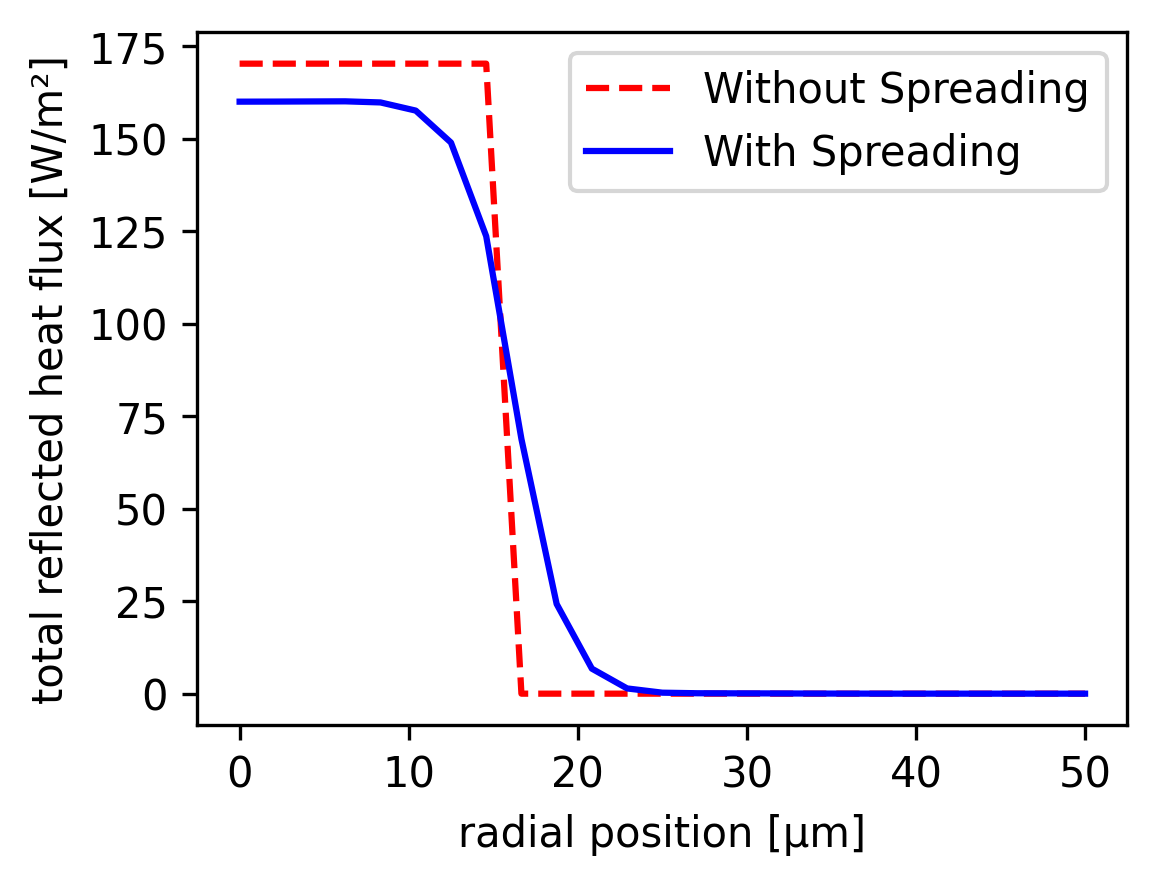}
    \caption{for step function gradient, w/o spread is 6.4\% greater at \(r=0\)}
    \label{fig:SphReflectedHeatFluxStepFunc_1cm}
\end{subfigure}%
\begin{subfigure}{.49\textwidth}
    \centering
    \includegraphics[width=\textwidth]{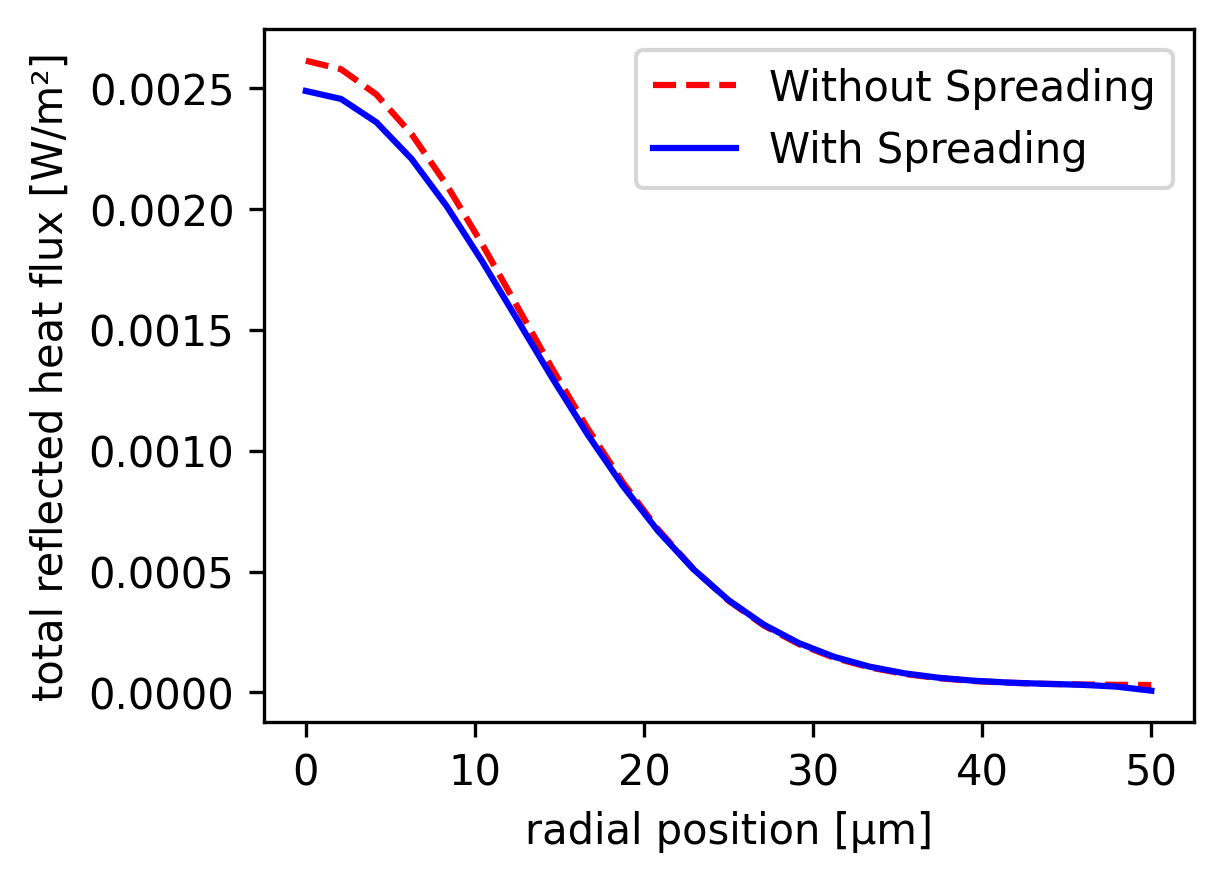}
    \caption{for Gaussian gradient, w/o spread is 5.0\% greater at \(r=0\)}
    \label{fig:SphReflectedHeatFluxGaussian_1cm}
\end{subfigure}
\caption{Total reflected heat fluxes with spreading and without spreading for the spherical coordinate system potential space, leading edge radius of \SI{1}{\centi\meter}, and quasi-neutral plasma density of \(2.42 \times 10^{19}\;\text{m}^{-3}\).}
\label{fig:SphREMModelReflectedHeatFluxes_1cm}
\end{figure}

\begin{figure}[H]
\centering
\begin{subfigure}{.49\textwidth}
    \centering
    \includegraphics[width=\textwidth]{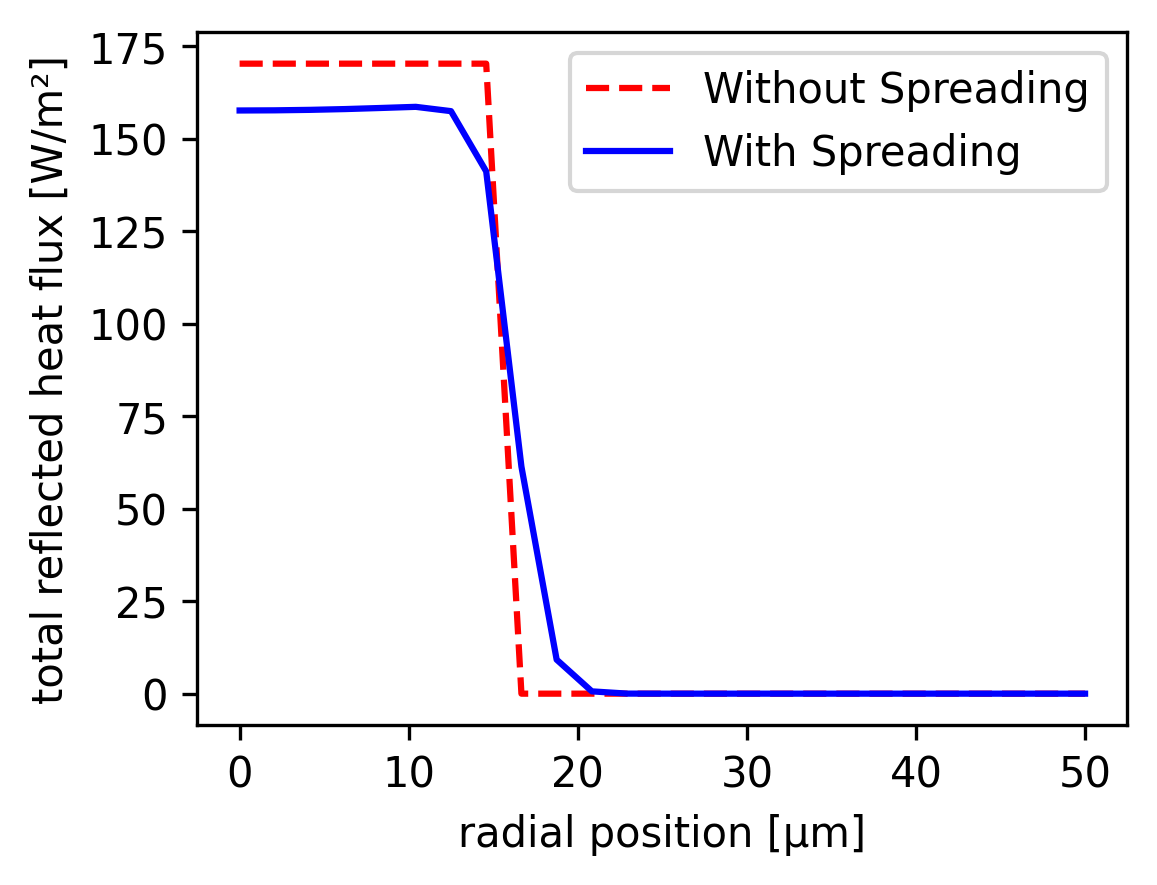}
    \caption{for step function gradient, w/o spread is 8.0\% greater at \(r=0\)}
    \label{fig:SphReflectedHeatFluxStepFunc_1mm}
\end{subfigure}
\begin{subfigure}{.49\textwidth}
    \centering
    \includegraphics[width=\textwidth]{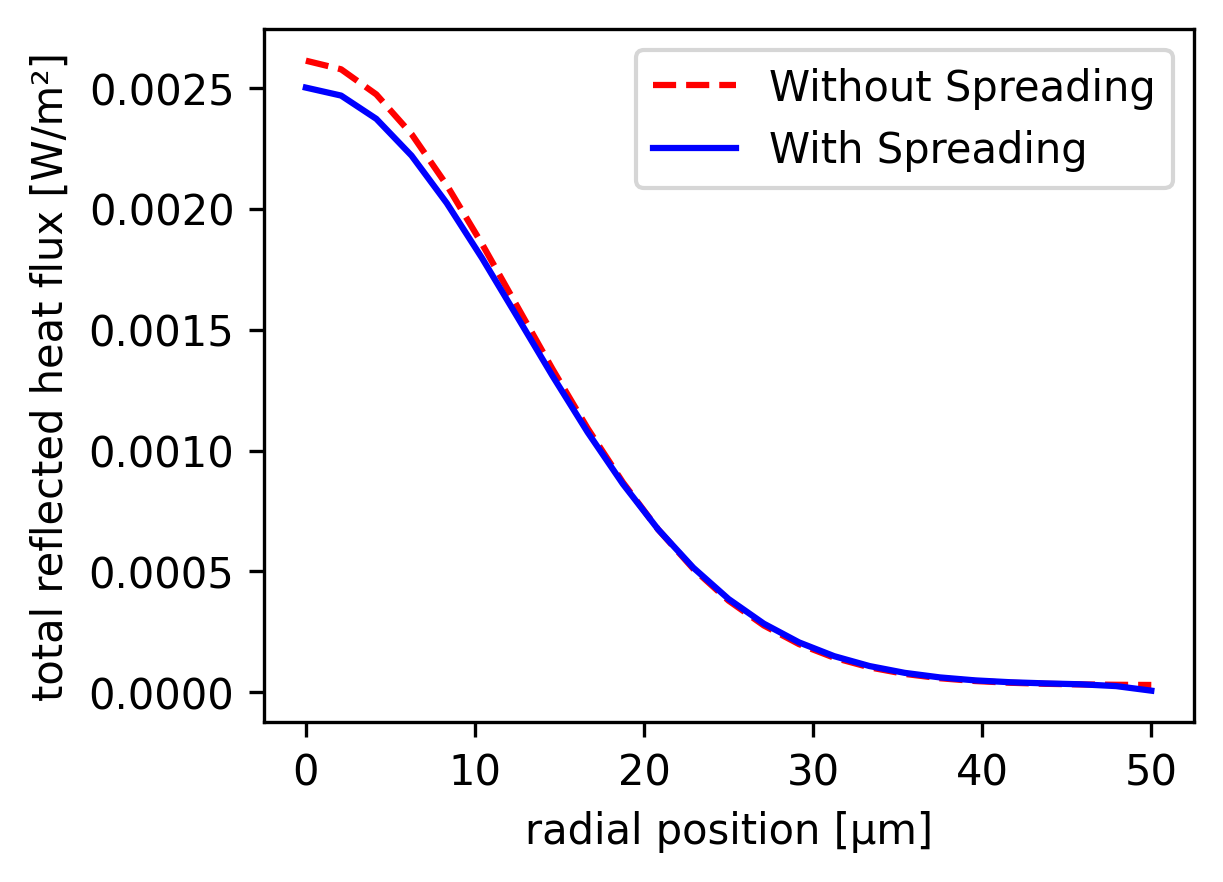}
    \caption{for Gaussian gradient, w/o spread is 4.5\% greater at \(r=0\)}
    \label{fig:SphReflectedHeatFluxGaussian_1mm}
\end{subfigure}
\caption{Total reflected heat fluxes with spreading and without spreading for the spherical coordinate system potential space, leading edge radius of \SI{1}{\milli\meter}, and quasi-neutral plasma density of \(2.42 \times 10^{21}\;\text{m}^{-3}\).}
\label{fig:SphREMModelReflectedHeatFluxes_1mm}
\end{figure}

To evaluate heat spreading utilizing our methodology, we construct two different temperature gradients in the radial direction to apply to the reduced mesh, a step function  profile, and a Gaussian profile, pictured below in Fig. \ref{fig:PrescribedTempGradients}. The red points signify the temperatures at which the electron velocity PDF Eq. \eqref{eq:REMVelocityDistFunc} is computed, as linear interpolation of the electron velocity PDF is performed between those temperatures. The number of temperatures at which we compute the electron velocity PDF is intentionally minimized as much as possible while still maintaining a decent approximation of the prescribed thermal gradient, as computing each electron velocity PDF is computationally expensive.
\begin{figure}[H]
\centering
\begin{subfigure}{.49\textwidth}
    \centering
    \includegraphics[width=\textwidth]{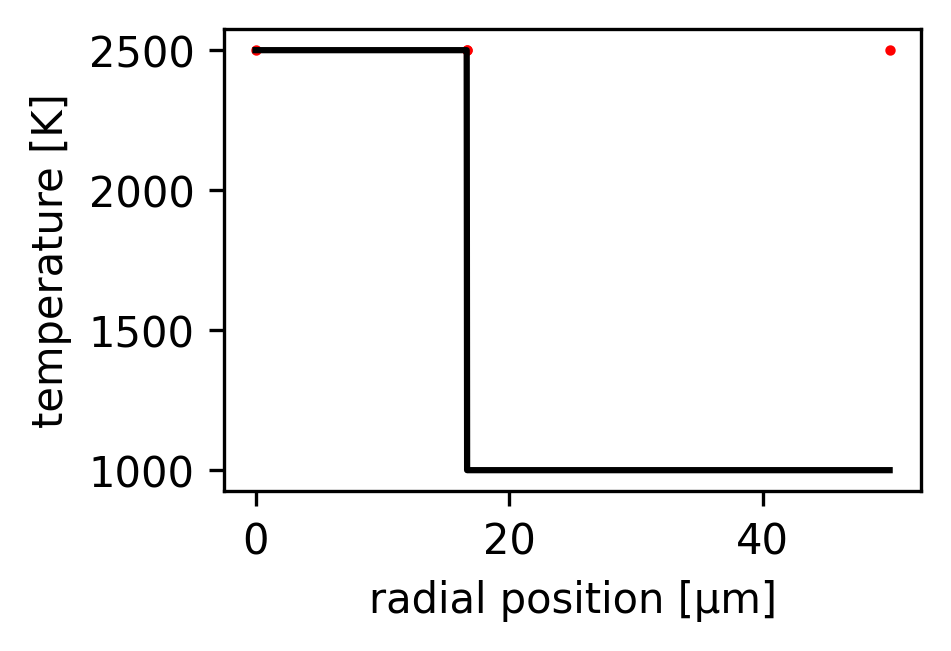}
    \caption{step function gradient}
    \label{fig:PrescribedTempGradientStepFunc}
\end{subfigure}%
\begin{subfigure}{.49\textwidth}
    \centering
    \includegraphics[width=\textwidth]{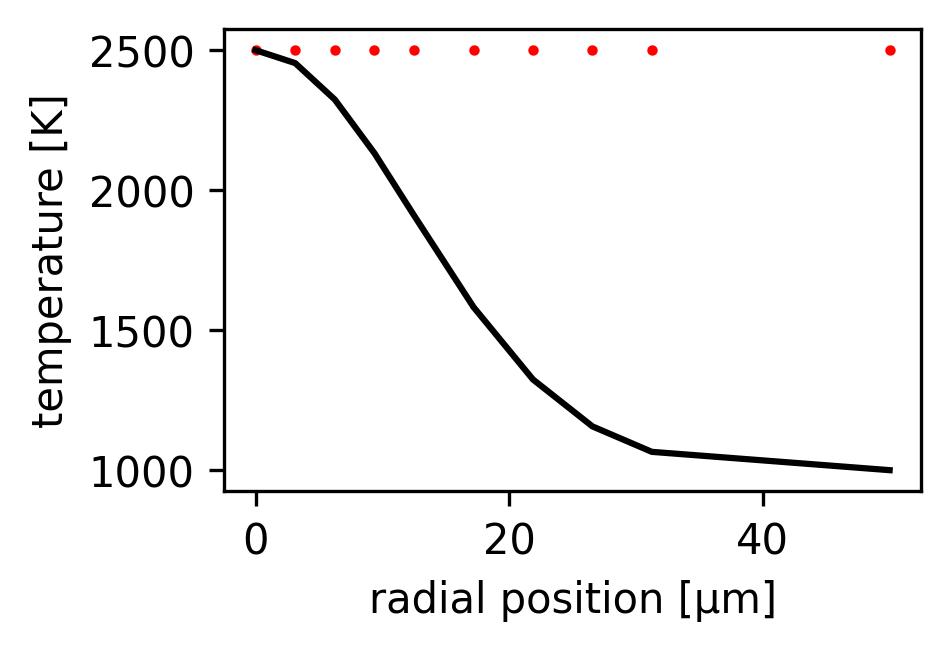}
    \caption{Gaussian gradient}
    \label{fig:PrescribeTempGradientGaussian}
\end{subfigure}
\caption{Prescribed temperature gradients on sample surface in radial direction.}
\label{fig:PrescribedTempGradients}
\end{figure}
Applying Eqs. \eqref{eq:SpreadingFluxReflected} and \eqref{eq:NonspreadingFluxReflected} to the two temperature gradients gives the following results in Figs. \ref{fig:CarREMModelReflectedHeatFluxes_1cm}-\ref{fig:SphREMModelReflectedHeatFluxes_1mm}. As a sanity check, we note that the total reflected energies calculated by integrating the heat flux times \(2\pi r \, dr\) match for the with-spreading and without-spreading cases. First, it is immediately apparent that the total reflected flux near the sample center \(r = 0\) is greater for the without-spreading case than in the with-spreading case, as expected, because we allow electrons to move from the hot to cold region. And intuitively, the shape of the total reflected flux as a function of position mirrors the shape of the underlying temperature gradient.
Specifically, for all three potential cases, for the step function temperature gradient, we observe a reduction and broadening of the total reflected heat flux, as electrons from the hot region between 0 and \SI{20}{\micro\meter} travel to the cold region. In all the cases for the Gaussian temperature gradient, we also observe a slight reduction in the total reflected heat flux at radial position \(r = 0\), although the broadening is not as pronounced due to the temperature gradient being less steep. Consistent among the results of all three coordinate systems, we notice that the percentage by which the total reflected heat flux at the stagnation point \(r = 0\) without spreading is greater than that with spreading is at least double in the case of the step function temperature gradient, compared to that of the Gaussian gradient. With regard to the Cartesian results specifically, the percentage by which the without-spreading flux is greater than with-spreading flux at \(r = 0\) increases for the step function temperature gradient as we decrease the LE radius from \SI{1}{\centi\meter} to \SI{1}{\milli\meter}. And this percentage decreases for the Gaussian temperature gradient for the same decrease in LE radius.
But for the cylindrical results, the percentage by which without-spreading flux is greater than with-spreading flux at \(r = 0\) \textit{decreases} for the step function temperature gradient as we decrease the LE radius, while the percentage \textit{increases} for the Gaussian temperature gradient given the same decrease in LE radius. Now in the spherical results, the aforementioned percentage increases for the step function gradient and decreases for the Gaussian temperature gradient, like it does for the Cartesian results. Also for the spherical results, the changes in the percentage by which the without-spreading flux is greater than the with-spreading flux as the LE radius decreases, are slightly smaller than those in the cylindrical results, which is likely due in part to the larger virtual cathode barrier present in the spherical coordinates potential space. 
Also, because the virtual cathode magnitudes for the three potential spaces are all slightly different, we find that a deeper virtual cathode yields a higher peak reflected heat flux at \(r = 0\), which is anticipated as more electrons should reflect due to seeing a higher potential barrier. Another factor that contributes to the phenomena discussed, is that the cylindrical and spherical potential spaces extend 25 to 30 microns from the wall, whereas the Cartesian potential space has only a size of 21 microns. Therefore, not only do the cylindrical and spherical coordinate potential spaces have slightly smaller virtual cathode barriers than that of the Cartesian potential, but they also facilitate a potentially larger electron travel distance due to their VC positions and shapes. Perhaps the most notable observation, is that the percentage difference between the with-spreading and without-spreading cases is of the same order of magnitude for both leading edge radii tested, given that the spreading is computed for the same \SI{50}{\micro\meter} radius mesh in Fig. \ref{fig:SpreadingReducedMesh}; this implies that heat spreading impacts a greater percentage of surface area for a smaller leading edge radius, therefore this effect should be accounted for as leading edge radii are engineered to smaller and smaller radii.

\section{Conclusion}
Using the random energy model of McCarthy et al. \cite{McCarthy2014} we have developed a numerical model of thermionic and photoemission cooling and heat spreading, and examined how electrons moving laterally under the acceleration of a potential sheath ``spread out'' reflected heat flux. While the reduction of total reflected heat flux due to heat spreading at the stagnation point \(r = 0\) is under 10\% for the potential spaces and temperature gradients tested, it may be possible to observe heat spreading over longer electron travel distances, with engineered potentials and engineered incident light characteristics. Furthermore, because we observed similar reflected heat fluxes for both leading edge radii, we would like to quantify to what extent the percentage of surface area experiencing heat spreading increases for progressively smaller leading edge radii. Future work will also consider how prescribed potential space, temperature gradient, incident light frequency range, and light intensity as a function of frequency affect reflected heat fluxes and the relevant length scale of heat spreading, through parametric studies.

\end{document}